\documentclass[letterpaper,12pt]{article}

\usepackage{graphicx}     
\usepackage{epsf}

\begin{document}

\title{Asymmetric  Coupling in  Two-Channel Simple Exclusion Processes}
\author{Ekaterina Pronina \and Anatoly B. Kolomeisky\thanks{tolya@rice.edu} \\
Department of Chemistry, Rice University, Houston, TX 77005}

\maketitle

\begin{abstract}

Simple exclusion processes for particles  moving along two parallel lattices and jumping between them  are theoretically investigated  for asymmetric rates of transition between the channels. An approximate theoretical approach, that describes the particle dynamics exactly in any vertical cluster of two parallel sites and neglects the correlations between the different vertical clusters, is applied to calculate  stationary-state density profiles, currents and  phase diagrams. Surprisingly, it is found that  asymmetry in the coupling between the channels leads to a very complex phase  behavior that is very different from two-channel simple exclusion processes with  symmetric coupling. There are seven stationary-state phases in the simple exclusion processes with asymmetric transition rates  between the channels, in contrast to three phases found for the systems with symmetric coupling. In addition, a new maximal-current phase with a domain wall in the middle of the lattices, that has no analogs in other exclusion processes, is observed. Although the explicit calculations are presented only for the case of full asymmetry, when the particles can only jump between the channels in one direction, the properties of two-channel simple exclusion systems with general asymmetry are also discussed. Theoretical predictions are in excellent agreement with extensive computer Monte Carlo simulations.

\end{abstract}

\pagebreak

\section{Introduction}

Asymmetric simple exclusion processes (ASEP) play a critical role for understanding multiple non-equilibrium phenomena in chemistry, physics and biology \cite{derrida98,schutz}. ASEPs have been extensively studied and applied for description of kinetics of biopolymerization \cite{macDonald68}, protein synthesis \cite{shaw03,chou04}, transport of motor proteins in biological cells \cite{klumpp03}, polymer dynamics in dense medium \cite{evans94}, car traffic processes \cite{nagel96} and modeling of ant trails \cite{chowdhury04}. 

The majority of studied asymmetric exclusion models analyze the multi-particle dynamics along a single lane \cite{derrida98,schutz}. At the same time, the more realistic description of many processes, such as car traffic and the biological transport of motor proteins \cite{klumpp03,nagel96}, suggest that it is important to study  multi-lane  ASEPs. There have been several theoretical investigations of two-lane simple asymmetric exclusion processes \cite{popkov01,pronina04,pronina05,mitsudo05,harris05}. We introduced earlier  two-channel simple exclusion models where particles can move along the channels and between them \cite{pronina04}. Using a vertical cluster mean-field approach the stationary properties have been analyzed and compared with extensive computer Monte Carlo simulations. Only the case of symmetric coupling, when the particles jump with equal rates between the lanes, has been considered. Mitsudo and Hayakawa \cite{mitsudo05} extended these models to more general asymmetric rates of moving between the channels, and it was shown that kinks, or domain walls, in both channels synchronize their motion. However, the theoretical analysis has been based on the mean-field decoupling approximation that totally neglects all correlations in the system, and significant deviations from Monte Carlo computer simulations results have been observed, especially when the coupling is asymmetric. The effect of disorder in the transition rates and current-density relations for limited range of parameters have also been  studied  for two-lane  exclusion processes with asymmetric coupling \cite{harris05}. 

The goal of  this paper is to investigate the general stationary-state properties of two-channel exclusion processes with asymmetric coupling for all possible sets of parameters. We consider the effect of asymmetry in the transition rates between the lanes  by using the vertical cluster mean-field approach, that takes into account the correlations between the channels. In addition,  extensive computer Monte Carlo simulations are performed in order to test the theoretical predictions.

The paper is organized as follows. Theoretical description of the model and mean-field analysis are presented in Section 2. The results of Monte Carlo computer simulations and comparison with theoretical predictions are discussed in Section 3. Summary and conclusions are presented in the final Section 4.

\section{Theoretical Description}

\subsection{Model} \label{Model_Description}

We consider a system of two parallel one-dimensional lattices  where identical particles can move  along the channels and between them, as shown in Fig. 1. Each lattice has $L$ sites, and every site  can be occupied by no more than one particle or it can be empty. At every time step a site is randomly chosen from the lane 1 or 2. In the bulk of the system the particle dynamics is described by the following rules. A particle  can change the lane  with the rate $w_{1}$ or $w_{2}$ from the channel 1 or 2, correspondingly, if the vertical neighboring site is available - see Fig. 1. The particle at site $1 \le i < L$ can also move from left to right along the same channel to site $i+1$, if this site is empty. The rates for horizontal moves from the site $i$ depend on the occupancy state of the site $i$ on another lane. The particle on lane 1  moves to the right with the rate $1-w_{1}$ if the vertical neighboring site is not occupied, otherwise it jumps with the rate 1. Similarly, for the particle  on lane 2  the horizontal transition rates are equal to $1-w_{2}$ (the vertical neighbor is empty) or 1. These rules satisfy the condition that the total probability per unit time of leaving  the site $i$ (in any direction) is always equal to 1 \cite{pronina04}. 

In addition, there are special entrance and exit dynamic rules at the boundaries. Particles can enter the system with the  rate $\alpha$ if any of the first sites at each lane is not occupied. When a particle reaches the exit site $L$ it can leave the system with the rate $\beta$ if the exit site on another lane is occupied. Otherwise, the exit rates are $\beta(1-w_{1})$ and $\beta(1-w_{2})$ for the channel 1 and 2, correspondingly.

\begin{figure}[h] 
\centering
\includegraphics[scale=0.6, clip=true]{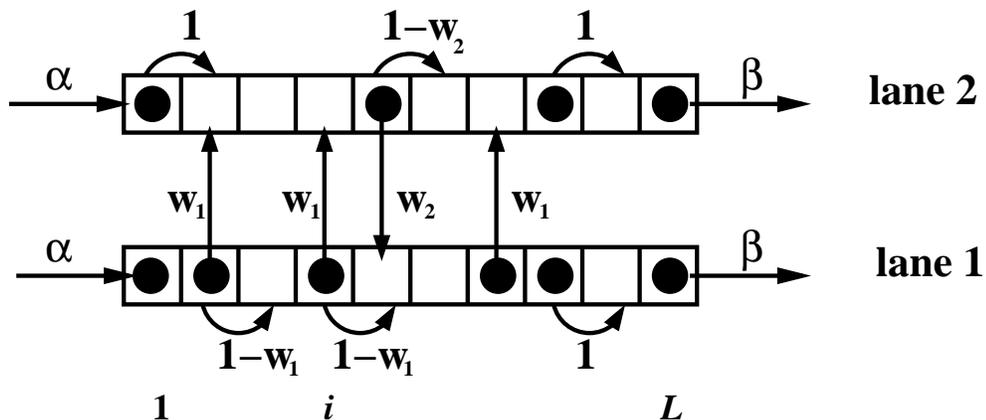}
\caption{Schematic view of the model for a two-channel ASEP with asymmetric coupling. Allowed transitions are shown by arrows. The inter-channel transition rates are equal to $w_{1}$ and $w_{2}$. The horizontal transition rates are $1-w_{1}$ and $1-w_{2}$ for the channel 1 and 2, correspondingly, if the vertical neighboring sites are empty. Otherwise, particle move with the rate 1. Entrance rates are equal to $\alpha$ at both lanes. Exit rates are equal to $\beta(1-w_{1})$ and $\beta(1-w_{2})$ for the lane 1 and 2, respectively, if the exit vertical cluster is half-occupied. For the fully-occupied vertical cluster at the exit the rate of leaving is $\beta$.}
\label{model}
\end{figure}

For simplicity, in the present work we consider only the case of full asymmetry in the vertical transition rates  with $w_{1}=1$ and  $w_{2}=0$. However, our analysis can also be straightforwardly applied for more general asymmetric couplings in two-channel ASEPs.  Note also that there is a particle-hole symmetry in the system, and it can be mapped into itself by exchanging everywhere the labels 1 and 2 and the rates $\alpha$ and $\beta$. This symmetry is important for understanding the stationary-state properties of two-lanes exclusion processes.

\subsection{Mean-field Analysis}

Theoretical study of two-channel ASEP with symmetric coupling \cite{pronina04} indicates that correlations between the channels strongly influence the stationary-state properties of the system, however the horizontal correlations inside the lanes are relatively weak. The effect of inter-channel correlations is even larger for two-channel ASEPs with asymmetric coupling \cite{mitsudo05}. This suggests that a cluster mean-field approach \cite{pronina04}, that explicitly takes into account the correlations inside the vertical cluster of lattice sites,  is the most  appropriate and convenient theoretical tool for analyzing these systems. 

The basic quantities of the vertical cluster mean-field approach are the probabilities to find any vertical cluster in one of four possible states, as shown in Fig. 2. Considering the lattice sites far away from the boundaries of the system, it is assumed that the occupation of vertical clusters is independent of the position along the channels. We define $P_{11}$ as a probability to find a vertical cluster with both lattice sites filled, $P_{10}$ and $P_{01}$ as  probabilities to have a half-empty vertical cluster with a particle at the channel 1 or 2, respectively, and  $P_{00}$ as a probability to have no particles at both lattice sites. The conservation of probability requires that
\begin{equation} \label{prob_norm}
P_{11}+P_{01}+P_{10}+P_{00}=1. 
\end{equation}
In addition, the bulk densities at each channel  can be expressed in terms of the vertical cluster probabilities,
\begin{equation}
\rho_{1}=P_{11}+P_{10}, \quad \rho_{2}=P_{11}+P_{01}.
\end{equation}

\begin{figure}[h] 
\centering
\includegraphics[scale=0.6, clip=true]{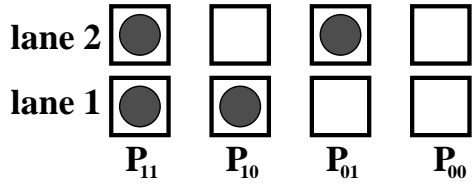}
\caption{Four different configurations for  vertical clusters of lattice sites. $P_{11}$, $P_{10}$, $P_{01}$ and $P_{00}$ are the corresponding  probabilities for each state.}
\label{cond_prob}
\end{figure}

The dynamics of the system  can be described via Master equations for evolution of every vertical cluster state. Specifically, for the fully occupied cluster we have
\begin{equation}
\frac{d P_{11}}{dt}=P_{11}P_{10}+P_{11}P_{01}+P_{01}P_{10}-2P_{11}P_{00}-P_{11}P_{10}-P_{11}P_{01}.  
\end{equation}
In the limit of $t \rightarrow \infty$ the system reaches a stationary state with $\frac{d P_{11}}{dt}=0 $, and this equation simplifies into
\begin{equation} \label{eq4}
P_{01}P_{10}=2P_{11}P_{00}.
\end{equation}
Similarly, for the half-filled vertical clusters it can be shown that
\begin{equation}
\frac{dP_{10}}{dt}=P_{11}P_{00}+P_{11}P_{00}+P_{11}P_{10}-P_{10},  
\end{equation}
that at the stationary-state limit  ($\frac{dP_{10}}{dt}=0$) reduces to
\begin{equation} \label{eq6}
P_{10}(1-P_{11})=2P_{11}P_{00}.
\end{equation}
From Eqs. (\ref{prob_norm}), (\ref{eq4}) and (\ref{eq6}) it is possible to calculate the stationary-state probabilities of the different vertical cluster states, and consequently all properties of two-channel ASEPs with asymmetric coupling can be obtained.

In the bulk of the system the stationary currents are given by the following expressions,
\begin{equation} \label{J_bulk}
J_{bulk,1}=P_{11}(1-P_{11}-P_{10}),  \quad  J_{bulk,2}= (P_{11}+P_{01})(1-P_{11}-P_{01}).  
\end{equation}
The particle currents at the boundaries are different form the bulk expressions. For entrance we obtain
\begin{equation} \label{J_entr}
J_{entr,1}=\alpha(1-P_{11}-P_{10}),  \quad  J_{entr,2}=\alpha(1-P_{11}-P_{01});  
\end{equation}
while at the exit it can be shown that
\begin{equation} \label{J_exit}
J_{exit,1}=\beta P_{11},  \quad  J_{exit,2}=\beta(P_{11}+P_{01}).  
\end{equation}
Although the  currents on the individual lanes can differ from each other, the overall current of the system at the stationary state is always constant, 
\begin{equation} \label{J_total}
J_{total}= J_{bulk,1}+J_{bulk,2}=J_{entr,1}+J_{entr,2}= J_{exit,1}+J_{exit,2}.  
\end{equation}

Solving together Eqs. (\ref{prob_norm}), (\ref{eq4}) and (\ref{eq6}) leads to a conclusion that at large times there are three possible cases. In one region of phase space for all bulk lattice sites we have
\begin{equation} \label{case1}
P_{10}=P_{11}=0,
\end{equation}
which means that the bulk density at the lane 1 is zero. In another region of the parameter's space 
\begin{equation} \label{case2}
P_{10}=P_{00}=0,
\end{equation}
that leads to $\rho_{2}=1$, i.e., the channel 2 is fully occupied by the particles.  There is also a case when the two-channel system can be divided in two parts, with Eq. (\ref{case1}) valid in one part while Eq. (\ref{case2}) is satisfied in another part.

Let us consider first the region in the parameter's space when the system is described by  Eq. (\ref{case1}). In this case there are only two possible states for the vertical clusters, the cluster with both empty sites  (00) and the cluster with the occupied site on lane 2 (01).  The expressions for the particle currents are significantly simplified, 
\begin{eqnarray}\label{currents1}
J_{bulk,1}=0, &  J_{bulk,2}= P_{01}(1-P_{01}), \nonumber \\
J_{entr,1}=\alpha,  &  J_{entr,2}=\alpha(1-P_{01}) \nonumber \\ 
J_{exit,1}=0,  &  J_{exit,2}=\beta P_{01}.
\end{eqnarray}
Then the  dynamics of the two-lane system can be viewed as an effective one-channel transport  with ``particles'' given by (01) vertical clusters and ``holes'' being (00) vertical clusters. The particles are entering into the system with an effective rate $\alpha_{eff} \ne \alpha$, that can be calculated from Eqs. (\ref{currents1}). This is due to the fact that there are two entrance currents. At the same time  the effective particles are exiting with the rate $\beta$ since there is only one exit current. It is known \cite{derrida98,schutz} that the corresponding one-channel ASEP has low-density, high-density and maximal-current phases, and we conclude that there are also three phases, labeled as (0,LD), (0,HD) and (0,MC), in this part of the phase space for the two-channel ASEP with asymmetric coupling. The indexes in these labels for the different phases reflect the fact that $\rho_{1}=0$ is always  in the bulk of the system, while the state of the channel 2 depends on the entrance and exit rates. 

The phase boundaries and the effective entrance rate  $\alpha_{eff}$ can be determined in the following way. In the low-density phase (0,LD) the entrance current determines the total current in the system, namely,
\begin{equation}
J_{entr,1} + J_{entr,2}= J_{bulk,1} + J_{bulk,2}.
\end{equation}
From Eqs. (\ref{currents1}) we have
\begin{equation}
\alpha+\alpha(1-P_{01})= \alpha_{eff} (1-P_{01})=P_{01}(1-P_{01}),
\end{equation}
with $P_{01}=\alpha_{eff}$. It can be shown  that
\begin{equation}
\alpha_{eff}=P_{01}=\rho_{2}=\frac{1+\alpha-\sqrt{(1+\alpha)^{2}-8\alpha}}{2}, \quad P_{00}=\frac{1-\alpha+\sqrt{(1+\alpha)^{2}-8\alpha}}{2}.
\end{equation}
This phase exists for $\alpha_{eff} < \beta$ and $\alpha_{eff} < 1/2$, that leads to the following  conditions, 
\begin{equation}
\beta > \frac{1+\alpha-\sqrt{(1+\alpha)^{2}-8\alpha}}{2}, \quad \alpha < 1/6,
\end{equation}
with the total current in this phase given by
\begin{equation}
J_{total}=\alpha_{eff}(1-\alpha_{eff})=(\alpha/2) \left[3-\alpha + \sqrt{(1+\alpha)^{2}-8\alpha} \right].
\end{equation}
In the phase (0,HD) the exit processes determine the  particle dynamics in the channel 2, and this phase exists for $\beta <\alpha_{eff}$ and $\beta < 1/2$. It can be easily calculated that in this phase 
\begin{equation}
P_{01}=\rho_{2}=1-\beta, \quad P_{00}=\beta, \quad J_{total}=\beta(1-\beta).
\end{equation}
The conditions $\alpha_{eff} > 1/2$ and $\beta >1/2$ specify the phase (0,MD) with the maximal current at the channel 2. The stationary-state  properties of this phase are given by
\begin{equation}
P_{01}=P_{00}=1/2, \quad J_{total}=1/4.
\end{equation}
It is also important  to note that because the total particle current through the system cannot be larger than 1/4 all three phases that satisfy the Eq. (\ref{case1}) cannot exist for $\alpha > 1/2$.

Similar calculations can be performed for the region of the phase space where Eq. (\ref{case2}) is valid. Since in this case only the fully filled (11) and half-filed (01) vertical clusters can exist in the bulk, that leads to $\rho_{2}=1$ and the system again can be mapped into effective one-channel ASEP. Then there are three possible phases, called (LD,1), (HD,1) and (MC,1), can be found for this range of parameters. The particle currents are given by
\begin{eqnarray}\label{currents2}
J_{bulk,1}=P_{11}(1-P_{11}), &  J_{bulk,2}=0, \nonumber \\
J_{entr,1}=\alpha (1-P_{11}),  &  J_{entr,2}=0, \nonumber \\ 
J_{exit,1}=\beta P_{11},  &  J_{exit,2}=\beta .
\end{eqnarray}
The (11) vertical cluster play a role of new effective particles that enter the system with the rate $\alpha$ and exit with an effective rate $\beta_{eff}$. From Eqs. (\ref{currents2}) it can be shown that
\begin{equation}
\beta_{eff}=\frac{1+\beta-\sqrt{(1+\beta)^{2}-8\beta}}{2}.
\end{equation}
The phase (LD,1) exists for $\alpha < \beta_{eff}$ and $\alpha < 1/2$. For this phase we obtain
\begin{equation}
P_{11}=\rho_{1}=\alpha, \quad P_{01}=1-\alpha, \quad J_{total}=\alpha(1-\alpha).
\end{equation}
The conditions for the (HD,1) phase are given by $\beta_{eff} < \alpha$ and $\beta_{eff} < 1/2$, that leads to
\begin{equation}
\alpha > \frac{1+\beta-\sqrt{(1+\beta)^{2}-8\beta}}{2}, \quad \beta < 1/6.
\end{equation}
In this phase the stationary-state current is equal to
\begin{equation}
J_{total}=\beta_{eff}(1-\beta_{eff})=(\beta/2) \left[3-\beta + \sqrt{(1+\beta)^{2}-8\beta} \right],
\end{equation}
while the densities are
\begin{equation}
P_{11}=\rho_{1}=\frac{1-\beta+\sqrt{(1+\beta)^{2}-8\beta}}{2}, \quad P_{01}=\frac{1+\beta - \sqrt{(1+\beta)^{2}-8\beta}}{2}.
\end{equation}
The third possible phase is (MC,1),  defined for $\beta_{eff} > 1/2$ and $\alpha >1/2$. In this phase  the steady-state properties are given by
\begin{equation}
P_{01}=P_{11}=1/2, \quad J_{total}=1/4.
\end{equation}
Using the same arguments as above, all three phases described by Eq. (\ref{case2}) cannot be found for $\beta >1/2$. Note that these results can also be obtained by utilizing the particle-hole symmetry of the two-channel ASEPs with asymmetric coupling. The symmetry arguments  also suggest that $\alpha=\beta$ is a phase boundary between (0,HD) and (LD,1) phases.

Finally, for $\alpha > 1/2$ and $\beta > 1/2$ we have a situation when in one part of the two-channel system Eq. (\ref{case1}) is valid, while in the other part Eq. (\ref{case2}) determines the stationary behavior. We call this phase (MC,MC), and the total current in the system is equal to 1/4. Because of the particle-hole symmetry the boundary between  two different parts is expected to be found exactly at the middle of the lanes. 

Thus, the vector cluster mean-field analysis suggests that there are seven stationary phases in two-channel exclusion processes with asymmetric coupling. The predicted phase diagram is  shown in Fig. 3. There are two types of phase transitions can be observed in the system. The first-order phase transitions (shown by solid lines in Fig. 3) involve  a jump in the particle densities, while in the continuum phase transitions (dashed lines in Fig. 3) there are smooth changes in the density profiles. 

\begin{figure}[h] 
\centering
\includegraphics[scale=0.4, clip=true]{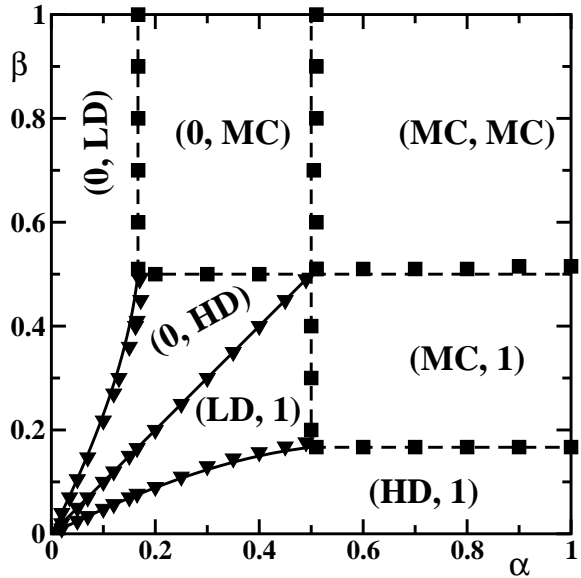}
\caption{ Phase diagram for the two-channel ASEP with full asymmetric inter-channel coupling ($w_{1}=1$ and $w_{2}=0$). Lines correspond to theoretical predictions, symbols are from the computer Monte Carlo simulations. Solid lines describe the first-order phase transitions, while dashed lines represent the continuum phase transitions.}
\label{diagram}
\end{figure}

Although our calculations are performed only for the case of full asymmetry in inter-channel coupling, similar calculations can be made for  systems with general asymmetry ($w_{1} > w_{2} >0$). It is expected to also find  seven stationary phases in the phase diagram, although the  phase boundaries, currents and the density profiles will depend explicitly on the vertical transition rates.

\section{Monte-Carlo Simulations and Discussions} \label{MC_sim}

Our theoretical approach based on the vertical cluster mean-field description of two-channel ASEP with asymmetric coupling predicts a complex phase diagram with many non-equilibrium phase transitions. It also allows to calculate stationary currents and density profiles in both channels. In order to test our theoretical results, a series of extensive Monte Carlo computer simulations have been carried out.

Since the predictions of the mean-field treatment are valid only in the thermodynamic limit, i.e., $L \rightarrow \infty$, our simulations has been performed for different lattice sizes with $L=100$, 500, 1000 and 5000. Most of our computer simulations results are obtained for $L=1000$ for which it was found that finite-size effects can be neglected. For faster computations the so-called BKL algorithm \cite{BKL} has been used. This method skips uneventful Monte-Carlo steps and it makes a corresponding correction in the calculation of the effective Monte Carlo time. Employing the  BKL algorithm has been especially useful for the cases of phase coexistence and for slow dynamics for weak entrance and exit rates. The density profiles and currents in our simulations have been computed by averaging $10^{7}$ Monte Carlo steps per site, although at the phase boundaries to obtain more accurate description  we used $10^{8}-10^{9}$ steps per site. In order to be confident that the system reached the steady state, typically first 3-5 $\%$ of all Monte Carlo steps have been neglected.  

The phase diagram computed from Monte Carlo simulations is presented in Fig. 3, and it can be seen that our theoretical predictions are in excellent agreement with computer simulations results. More information can be extracted from the density profiles that are shown in Fig. 4. Theoretically calculated  density profiles in both channels also agree quite well with the results from computer simulations, although there are deviations near the boundaries of the channels. This is due to the fact that in our approximate theoretical treatment it was assumed that probabilities of vertical clusters are independent of the position along the lattice. Obviously, this assumption is valid only in the bulk of the system far away from the boundaries. It should be noted that because of the mapping of two-channel ASEP with asymmetric coupling into effective one-channel exclusion process the densities near the entrance and exit can be calculated \cite{derrida98}.

\begin{figure}[tbp]
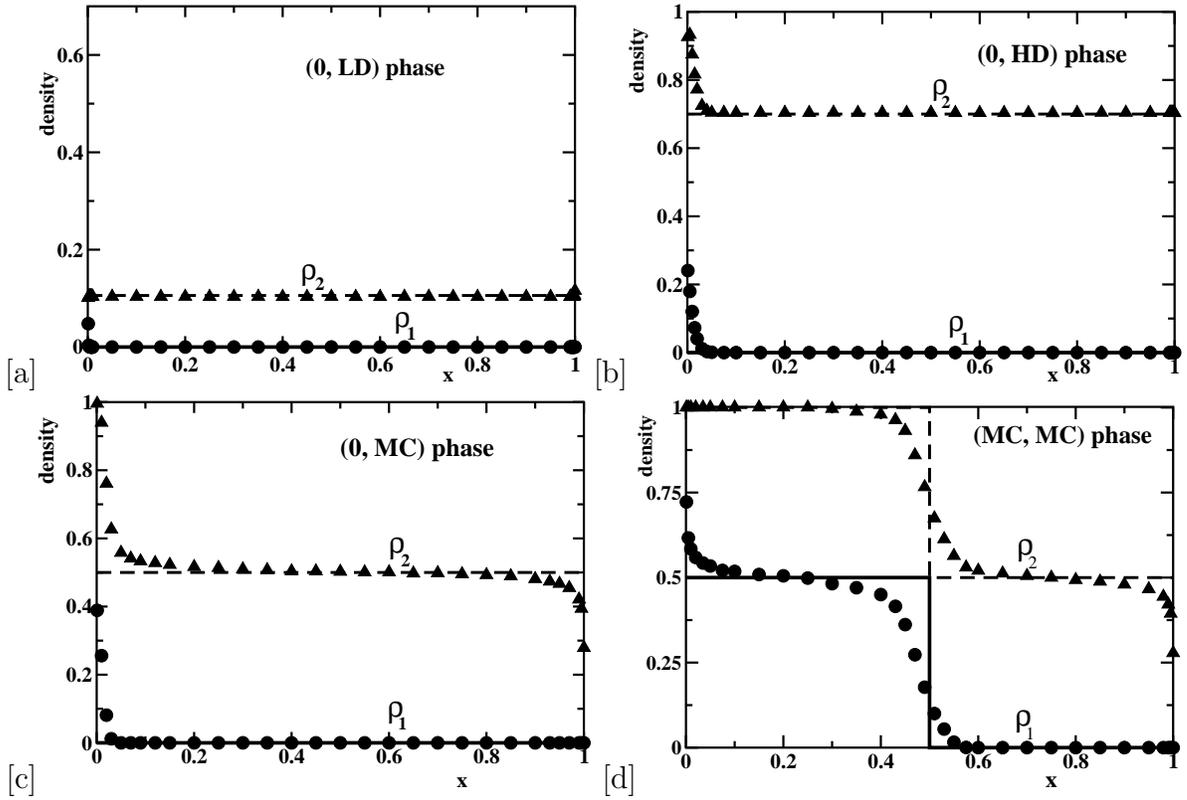

\centering
[a]\includegraphics[scale=0.3,clip=true]{Fig4a.eps}
[b]\includegraphics[scale=0.3,clip=true]{Fig4b.eps}
[c]\includegraphics[scale=0.3,clip=true]{Fig4c.eps}
[d]\includegraphics[scale=0.3,clip=true]{Fig4d.eps}
\caption{Density profiles for a) (0,LD) phase with $\alpha=0.05$ and $\beta=0.8$; b) (0,HD) phase with $\alpha=0.25$ and $\beta=0.3$; c) (0,MC) phase $\alpha=0.4$ and $\beta=0.9$; and d) (MC,MC) phase with $\alpha=\beta=0.9$. Symbols correspond to Monte Carlo computer simulations, lines describe the density profiles in each channel. Error bars, determined from standard deviations for simulations, are smaller than the size of the symbols.}\label{fig4}
\end{figure}

The analysis of Figs. 4a, 4b and 4c allows to understand the nature of (0,LD), (0,HD) and (0,MC) stationary phases. In all three phases  particles enter into the both channels (note that $\rho_{1} \ne 0$ at the left end of the system), but they exit only from the lane 2. Because of asymmetric inter-channel coupling the particles on the lane 2 have enough time to escape into the lane 1, and most of the particle transport in the system is taking place only along the channel 2.  The density profiles for  (LD,1), (HD,1) and (MC,1) phases can be easily obtained from (0,HD), (0,LD) and (0,MC) densities  by using the particle-hole symmetry arguments. In these three phases particles enter only through the channel 1 but there are two exiting currents. Because of the slow exiting processes and asymmetric coupling at large times the channel 2 is fully occupied by the particles, and the most of the particles  in the system move mostly  along the channel 1.

The most surprising result of our theoretical calculations is the existence of (MC,MC) phase shown in Fig. 4d. It has a domain wall, positioned in the middle of the system, that separates two different density profiles. In this phase particles enter the system only via the channel 1 but exit only through the channel 2, and there is an inter-channel flux of the particles near the middle position. The physical origin for this phase is the fact that  for $\alpha > 1/2$ and $\beta > 1/2$  the phases (0,MC) and (MC,1) can coexist together. Because of the particle-hole  symmetry the domain wall that separates two phases is exactly in the middle of the system. To the best of our knowledge, there are no such phases observed in any other driven diffusive systems \cite{derrida98,schutz}.

Our theoretical calculations  also  agree well with computer simulations for the density profiles at phase transitions  as shown in Fig. 5. Linear density profiles shown in Fig. 5 correspond to a first-order phase transition between (0,LD) and (HD,1) phases. The overall stationary dynamics and phase boundaries can also be understood by using the phenomenological domain wall description \cite{kolomeisky98}. 

\begin{figure}[tbp]
\centering
\includegraphics[scale=0.3,clip=true]{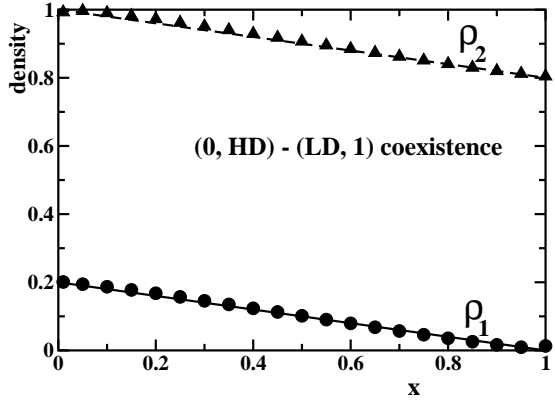}
\caption{Density profiles for phase transition line between (0,LD) and (HD,1) phase with $\alpha=\beta=0.2$. Symbols are  from Monte Carlo computer simulations, lines are from theoretical calculations.}\label{fig5}
\end{figure}

The dynamics of two-channel ASEP with asymmetric coupling can be well understood by analyzing the stationary currents presented in Fig. 6. Again our theoretical predictions match quite well the results from Monte Carlo computer simulations. For the fixed value of $\beta=0.6$ and for $\alpha <1/6$ the system is in (0,LD) phase and both entrance currents are increasing. At the phase transition ($\alpha=1/6$) between (0,LD) and (0,MC) phase $J_{entr,2}$ reaches the maximum and it starts to decrease for larger values of $\alpha$, while $J_{entr,1}$ is still growing. At $\alpha=0.5$ the second entrance current disappears and the first entrance current reaches a constant value of 1/4, and the system enters into (MC,MC) phase.

\begin{figure}[tbp]
\centering
\includegraphics[scale=0.3,clip=true]{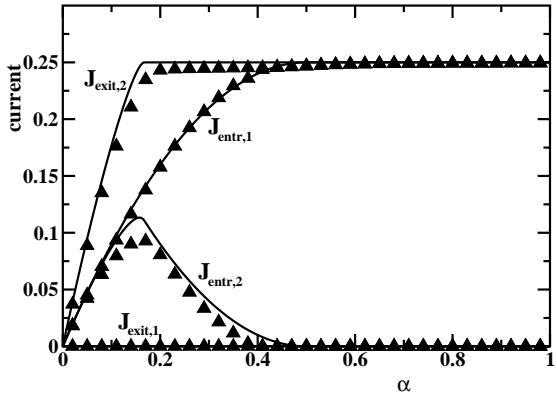}
\caption{Stationary particle currents for $\beta=0.6$. Symbols are  from Monte Carlo computer simulations, lines are from theoretical calculations.}\label{fig6}
\end{figure}

\section{Summary and Conclusions} \label{summary}

Two-lane simple exclusion processes with asymmetric inter-channel coupling have been investigated. To account for correlations between the channels, we utilized a mean-field method that allows to calculate exactly the probabilities of vertical clusters but neglects the correlations along the channels. It is shown that the asymmetry in inter-channel coupling produces a complex stationary-state behavior. There are seven stationary phases, and the particle dynamics in all phases can be understood in terms of effective one-channel ASEPs. The new (MC,MC)  phase  displays a domain wall between two density profiles in the middle of the system, and it has no analogs in other simple exclusion models. The specific position of phase boundaries depend on the degree of asymmetry, and for symmetric vertical transition rates the phase diagram simplifies with only three possible phases \cite{pronina04}.

There are several extensions of this model that will be important to investigate. In this paper the homogeneous vertical transitions rates have been considered. It will be interesting to understand the stationary properties of two-channel ASEP with inhomogeneous coupling. Another extension is to study the effect of adding the equilibrium Langmuir kinetics process of association and dissociation to one or to both lattices. Understanding these processes can help to describe better complex low-dimensional transport phenomena in chemistry, physics  and biology.

\section*{Acknowledgments}

The support  from the Welch Foundation (grant C-1559), the Alfred P. Sloan Foundation (grant BR-4418) and from the US National Science Foundation (grant CHE-0237105) is gratefully acknowledged. ABK also thanks T. Chou for valuable discussions.

\end{document}